# A first step towards an ecosystem meta-model for human-centered design in case of disabled users

Christophe Kolski [1][0000-0002-7881-6435], Nadine Vigouroux [2][0000-0003-4650-4362], Yohan Guerrier[1][0000-0003-1040-0799], Frédéric Vella[2][0000-0003-0761-9204] and Marine Guffroy[3][0000-0002-4954-3378]

[1] Univ. Polytechnique Hauts-de-France, CNRS, UMR 8201 – LAMIH, F-59313 Valenciennes, France
`{firstname.lastname}@uphf.fr`
[2] IRIT, UMR CNRS 5505, Université Toulouse 3, Toulouse, France
`{firstname.lastname}@irit.fr`
[3] Wiztivi, Carquefou, France
`marine.guffroy@wiztivi.com`

**Abstract.** The involvement of the ecosystem or social environment of the disabled user is considered as very useful and even essential for the human-centered design of assistive technologies. In the era of model-based approaches, the modeling of the ecosystem is therefore to be considered. The first version of a meta-model of ecosystem is proposed. It is illustrated through a first case study. It concerns a project aiming at a communication aid for people with cerebral palsy. A conclusion and research perspectives end this paper.

## 1 Introduction

For many years, different models have progressively taken a more and more important place, and even become essential, in the field of Human-Computer Interaction, with the objective of designing interactive systems. It is possible to quote on this subject, without concern of exhaustiveness, but rather of representativeness: the task model [14], the user model [16], the architecture model [4], the context model [9] [13], the adaptation model [29] or the dialogue model [7], etc. For our part, our research concerns the field of disability, and in particular the human-centered design of assistive technologies.

According to the International Classification of Functioning, Disability and Health [31], the interactions between environmental factors (physical, social and attitudinal) and personal factors must be taken into account. Indeed, these factors influence the limitations of activity and social participation of people with disabilities for the design of assistive technologies. In this paper, we focus on the social environment of people with disabilities. This is referred to as the ecosystem by [25] [26]. This ecosystem is composed for example of family and/or professional caregivers, friends, therapists or colleagues, and this in relation to a set of activities of the person with a disability. Our motivations are the following: We propose that the ecosystem model is now one of the important models to consider. Such model could support a new generation of tools supporting the HCD (Human-Centered Design) process in disability domain. But in HCI and disability domains, there is a lack concerning ecosystem modeling. It is why we propose in this paper the first version of a meta-model of the ecosystem. As a meta-model, this one could be instanciated in the form of ecosystem models in projects.

After the presentation of the background, a first step towards a meta-model of the ecosystem is proposed. Then, we propose a first instantiation of the meta-model from a previous case study, using a reverse engineering approach. A discussion, a conclusion and research perspectives end the paper.

## 2    Background

This section concerns the implication of members of the ecosystem of disabled people in human-centered processes, with a view of ecosystem modelling and meta-modelling. Human-centered design (HCD) is based on ISO 9241-210 [28], which defines steps for its implementation: 1) Understanding and specifying the context of use; 2) Specifying user requirements; 3) Prototyping solutions to meet user needs; 4) Evaluating the solutions. In this approach, the end users, as well as various other stakeholders, are at the heart of the above-mentioned steps. However, these activities encounter implementation constraints for the design of assistive technologies dedicated to people with disabilities (motor impairments, cognitive impairments, communication disorders, etc.) [2]. Their participation in HCD activities may be differentiated according to their impairments. Many methods are available, allowing to involve them, with or without members of their ecosystem [5].

Many studies are interested in this question of the direct or indirect participation of ecosystem members of disabled people. Several participatory design approaches involve ecosystem members in the design, such as [17], or in both design and evaluation, for example [30]. In general, a so-called co-design approach can involve also members of the ecosystem, see for example [1]. Other authors position their approach more globally according to HCD and involve also members of the ecosystem in the design and evaluation, such as [3] [6] [8] [12]. For instance, in their study [12], Barros and colleagues report that key stakeholders (family and professional caregivers, physicians, physical therapists, and Parkinson's patients) were involved throughout the system design process of the target system; this one concerned the self-management of Parkinson's disease.

More generally, the literature shows that the ecosystem actors involved in human-centered design, participatory design, co-design…, are most often relatives (family and professional caregivers), associated therapists depending on the impairment (speech therapist, physiotherapist, occupational therapist, etc.) and people with professional expertise (doctor, specialized educators, teacher, etc.) depending on the assistive technology designed. Thus, there is a growing interest in the literature regarding the involvement of ecosystem members. Some papers suggest a first categorization of stakeholders in the projects. For example, Augusto and colleagues propose four types in their study [3]: Primary users (People with Down's Syndrome), Secondary users (Main carers), Tertiary users (Other system users), Other (Those interested in the system but no direct users). A global classification (represented as a tree) from a study concerning categories of medical device users is also available in [33], but it is not focused on disability domain. In general, we note that the papers available in the literature do not include a model or meta-model of the ecosystem.

In addition, it is important to note that several meta-models are available in HCI domain in general. They address various aspect of HCI design. They concern for instance context-aware adaptation of user interfaces [29], processes for highly supporting flexibility of user interfaces [10] or context-aware adaptation of mobile applications [15]. But, to our knowledge, none of them addresses more or less explicitly ecosystem considerations.

The involvement of the members of the ecosystem of the disabled person, in relation to the technical system, can be modeled according to the approach proposed below.

## 3    Proposal: Towards an Ecosystem Meta-Model

We studied three user-centered design projects involving disabled people with communication disorders [23]. For each project, we analyzed their ecosystem. We found that the same types of actors could systematically be identified, and therefore involved in the methodological approach. Consequently, our aim was to propose a meta-model synthesizing the modeling elements involved, in relation to the three assistive technologies targeted [23] [26]. When considering other technologies, this meta-model could certainly be adapted or enriched.

As expressed by its *meta-* prefix, a meta-model (in our case, an ecosystem meta-model) is an abstraction allowing to describe models (i.e. ecosystem models) [18]. Figure 1 presents our first proposal for a meta-model aimed at providing support for the representation of the ecosystem of a user in a disability situation.

This meta-model is composed of the disabled person, one or more actors that can be considered as primary and secondary users of the assistive technology, and then one or more actors without interaction with the system (as defined in [2]). The disabled person has characteristics such as the type of impairment (e.g. mental or physical), the possible assistive technologies allowing to help them in different contexts of daily life, and finally the needs resulting from their disability situation. A primary actor (called Actor_P) is a person who directly uses the system. Most of the time, it is primarily the disabled person but other people in their environment can be primary actors of the system, in relation to one or more of its functionalities. For example a teacher who will regularly use a functionality of description of the day's activities of children with disabilities referring to this description on an interactive device [24] or an occupational therapist who design the assistive technology with and for the person with disabilities [8]. Concerning the secondary actor (called Actor_S) of the system, present in the meta-model, this user can interact for example, with the system if the primary user is unable to do so because of their disability or a constraining psychological situation such as stress. This actor also has a set of attributes. So, the type is intended to define whether the actor is, for example, a home care worker, a family member, a medical person, a teacher, etc. This person may also express needs in terms of communication assistance to help the person with a disability interact with people, known or unknown, in their environment. The secondary actor has a direct relationship with the person with a disability, from a communication perspective. Finally, the proposed meta-model includes an actor (called Actor_N) who does not directly operate the system. This person is, for example, the recipient of messages formulated by the person with a disability. This actor is a person who does not necessarily know the primary user, and therefore may have difficulties to understand them. The disabled person interacts in this case with an actor of type Actor_N through the communication assistive technology, either by being assisted by an actor of type Actor_S, or with their assistive technology. This depends on the degree of the person's disability. In the following case study section, a first instantiation of the meta-model is provided and commented.

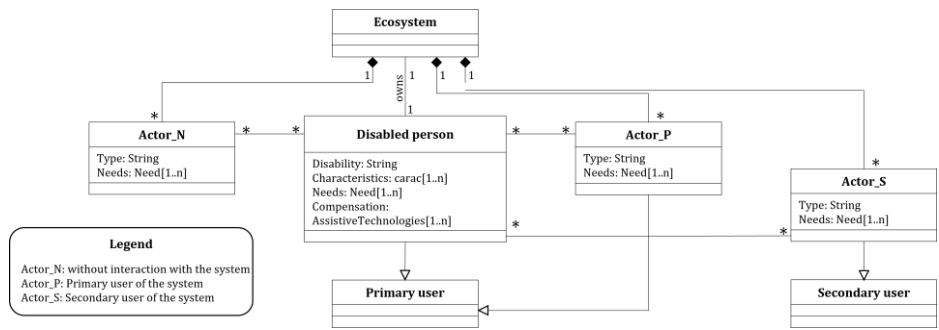

**Fig. 1.** Meta-model of ecosystem of user with disabilities.

## 4       First case study by Reverse Engineering

As described in Section 3, we studied three user-centered design projects involving disabled people and their ecosystem [23]. A meta-model resulting from this global analysis was proposed and described in Section 3. Using the principle of reverse engineering, it makes it possible to model the ecosystem a posteriori for each of these projects. This section illustrates a representative project.

This case study involved people with cerebral palsy. The project concerned the design, first prototyping and evaluation of a communication aid called ComMob (Communication and Mobility). After the description of the context, the ecosystem model related to this project is presented and commented. How the stakeholders in the ecosystem of the person with disabilities were involved in the HCD process is briefly presented. It is important to notice that the ecosystem model was built afterwards (by instantiation of the meta-model proposed), using a reverse engineering approach. It was possible to consult all the documents relating to the study and to interact with the members of the project team, with a focus on the role of the ecosystem in this project.

### 4.1 Context

In this case study, the target users have an athetosis-type cerebral palsy profile [32]. They make involuntary movements whose intensity can vary more or less strongly depending on the emotions. In addition, problems of precision in movements may make it difficult for them to use physical keyboards. These people also have speech problems due to dysarthria [11]. This means that the problems are usually in articulating words, but not in formulating correct sentences. Because of these different difficulties, people with this profile may require communication assistive technologies. A human-centered approach has been implemented with the aim of creating a communication assistive system for users with Cerebral Palsy [20]. In this system, called ComMob (Communication and Mobility), pictograms are organized in themes and categories. The user selects a set of pictograms to formulate sentences. The sentences are read by a text-to-speech system (see [19] for a description of ComMob).

### 4.2 Instantiation of the ecosystem meta-model

Figure 2 shows an instantiation of the meta-model proposed in Figure 1, relative to this case study. The disabled person has cerebral palsy and is the primary user. This means that this person directly uses this Communication tool. Then we have a set of caregivers, both accustomed and unaccustomed to communicating with people of this profile. These categories are part of the ecosystem of the disabled person. The familiar caregivers do not need to use ComMob because they understand the disabled person. However, in the case that the user is under severe stress, cannot speak or move, the regular caregivers can use ComMob instead of the disabled person, or help their to use it. These people are either professionals or relatives. The second category (unaccustomed caregivers) needs the tool to communicate with the disabled person. Often they are professionals. Other occasional interlocutors belong to the category of secondary users. This means that they need ComMob to communicate with the disabled person. In this last category, we find in particular passers-by, salesmen, drivers or health professionals.

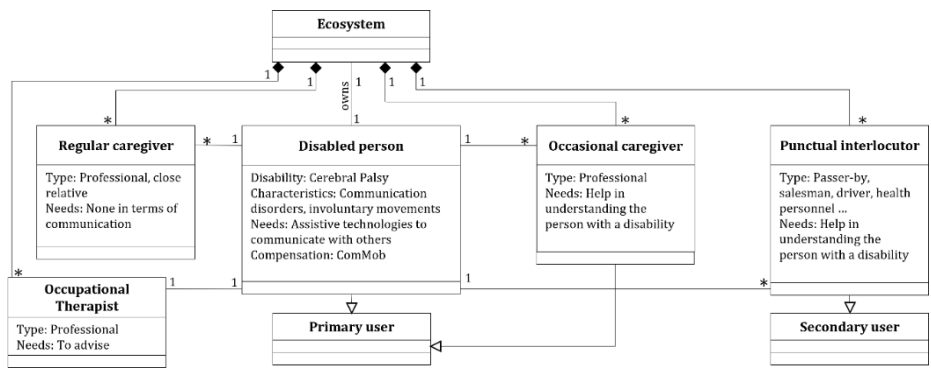

**Fig. 3.** Instantiation of the model in the case of ComMob project.

### 4.3 Actors' involvement in the HCD process

During the design of ComMob, the ecosystem was involved to gather the needs of future users. The disabled person, who is also the primary user, played a central role in the HCD process. Indeed, in the case of ComMob, the person with a disability was the designer of the application. Therefore, he had good knowledge in the field of cerebral palsy. In addition, he was in contact with an occupational therapist and other persons with the same disability. So he gathered information about the needs. A first mockup was made through his own needs, observations and testimonies obtained directly or on the internet. Then the designer conducted interviews with his family and his homecare providers to gather their opinions. Subsequently, tests were carried out with unaccustomed caregivers. For this purpose, the disabled person was in a public place. The task was to formulate a request to passers-by with ComMob [21]. Thanks to the data collected through observations and questionnaires filled in by passers-by, the project team was able to improve ComMob, for example by adding a button to repeat the previous sentence. Other tests were also performed in the laboratory. In one test, able-bodied participants used ComMob while simulating users with cerebral palsy, using a device manipulated by one of the evaluators. In this case, the ecosystem was not present, but an evaluator with cerebral palsy was an observer of the interactions performed by the participants [22].

## 5 Discussion

HCD has become a widely used design process over the years, resulting in a well-known standard. However, when the end users are disabled people with communication disorders, this model is no longer applicable and is no longer sufficient on its own. Indeed, the user cannot carry out alone the activities that the HCD provides. It is then necessary that actors of the ecosystem accompany the user. This paper proposes the first version of a meta-model to describe (by instantiation) such ecosystems.

The case study illustrates a first instantiation of this model, showing that these new actors have a place and a legitimate contribution in all the activities of the HCD. They have different professional expertise and social participation relationships with the disabled person. Furthermore, they may or may not be users of the assistive technology and may or may not assist the disabled person.

We note, however, that depending on the context in which the assistive technology will be used, the members of the ecosystem change. In a school context, the ecosystem will be composed of the educational team, while in a medical context it will be composed of health professionals. The implementation of HCD with the ecosystem requires some attention. The disabled user must not disappear from the design process [34], but must be involved by adapting the HCD tools or be represented by their ecosystem [27]. Indeed, the members of the ecosystem contribute and/or complete the collection of needs and accompany the user during the design and/or evaluation phases. The model may be very useful in highlighting the different actors involved in the project concerned (See also Section 6).

This ecosystem model, which takes the form of a UML class diagram, can be integrated into model-based approaches. Such a model can be linked to other models for design purposes. For example, through association or inheritance links with user models, it is possible to highlight or detail roles and characteristics. Connected to task models, it is possible to describe the tasks of each user category. Each context in which a user interacts with the assistive technology can be the subject of a context model. Each dialog associated with a human-machine interaction can be associated with a dialog model, and so on.

## 6 Conclusion

The software world is evolving quickly, with the advent of new model-based approaches. One observation is that models are playing an increasingly crucial role in projects. In parallel, the growing importance of the ecosystem in disability-related projects has led us to propose an ecosystem meta-model realized with the help of a UML class diagram. As a first validation, an illustration of the ecosystem model (again in the form of a class diagram), taking the form of instantiation of the meta-model, was provided. It comes from a case study concerning the ecosystem of people with cerebral palsy. Members of the ecosystem played important and diverse roles in the project.

The research perspectives are various. The proposal opens the way to the development of support tools for HCD approaches based on an ecosystem model or a library of reusable models. Moreover, it is necessary to analyze different projects, involving various disabilities, in order to highlight, for each of them, the corresponding ecosystem model. We have used the same modeling approach for other projects in which we have been involved; this work is not illustrated in this paper due to lack of space. We also plan to use it on new projects or projects where we have not been involved (provided the project documentation is available). The models obtained could lead to enrichments of the meta-model. As suggested in Section 5, we also plan to analyze the possible relationships between an ecosystem model (or the meta-model) and other types of models (user model, task model, context model, etc.). Another important perspective concerns the integration of the meta-model in model-driven engineering software environments and its implementation in various projects related to the disability domain.

The participation of the ecosystem in projects related to the field of disability is of growing importance. It should lead to the ecosystem model finding its place in the approaches and tools proposed by researchers and practitioners. The model could be associated with recommendations encouraging the systematic analysis of a set of possible categories of actor. It would support the study and choice of actors who could be involved in one or more phases of the user-centered design process [28]; for example, the family could be involved in the design and evaluation of mock-ups, while the doctor could be involved in the context understanding and specification phase, and so on. The model could also encourage reflection on the determination of primary (not necessarily only the disabled user) and secondary users. Indeed, depending on the characteristics and uses of the assistive technology, direct or indirect configuration or operation by an experienced or novice person may be necessary, depending on the context.